%% file: main.tex
\documentclass[sigconf]{acmart}

\def\BibTeX{{\rm B\kern-.05em{\sc i\kern-.025em b}\kern-.08emT\kern-.1667em\lower.7ex\hbox{E}\kern-.125emX}}

\copyrightyear{2020}
\acmYear{2020}
\setcopyright{acmcopyright}
\acmConference[ICPE '20]{Proceedings of the 2020 ACM/SPEC International Conference on Performance Engineering}{April 20--24, 2020}{Edmonton, AB, Canada} 
\acmBooktitle{Proceedings of the 2020 ACM/SPEC International Conference on Performance Engineering (ICPE '20), April 20--24, 2020, Edmonton, AB, Canada} 
\acmPrice{15.00}
\acmDOI{10.1145/3358960.3379126}
\acmISBN{978-1-4503-6991-6/20/04} 

\usepackage{balance}
\usepackage{soul}
\usepackage{color}

\begin{document}

\fancyhead{}

\title{Detecting Latency Degradation Patterns\\ in Service-based Systems}

\author{Vittorio Cortellessa}
\email{vittorio.cortellessa@univaq.it}
\author{Luca Traini}
\email{luca.traini@graduate.univaq.it}
\affiliation{
  \institution{University of L'Aquila}
  \country{Italy}
}

\begin{abstract}
Performance in heterogeneous service-based systems shows non-determistic trends.
Even for the same request type, latency may vary from one request to another.
These variations can occur due to several reasons on different levels of the software stack:
operating system, network, software libraries, application code or others.
Furthermore, a request may involve several Remote Procedure Calls (RPC),
where each call can be subject to performance variation.
Performance analysts inspect distributed traces and seek for recurrent patterns in trace attributes, 
such as RPCs execution time, in order to cluster traces in which variations may be induced by the same cause.
Clustering "similar" traces is a prerequisite for effective performance debugging.
Given the scale of the problem, such activity can be tedious and expensive.
In this paper, we present an automated approach that detects relevant RPCs execution time patterns associated to request latency degradation, i.e. \emph{latency degradation patterns}.
The presented approach is based on a genetic search algorithm driven by an information retrieval relevance metric and an optimized fitness evaluation.
Each latency degradation pattern identifies a cluster of requests subject to latency degradation with similar patterns in RPCs execution time.
We show on a microservice-based application case study that the proposed approach can effectively detect clusters identified by artificially injected latency degradation patterns.
Experimental results show that our approach outperforms in terms of F-score a state-of-art approach for latency profile analysis and widely popular machine learning clustering algorithms.
We also show how our approach can be easily extended to trace attributes other than RPC execution time (e.g. HTTP headers, execution node, etc.).  

\end{abstract}



\maketitle

\input{introduction2}

\input{problem}

\input{approach}
\input{evaluation}
\input{results}

\input{onestepahead}
\input{relatedwork}

\section{Conclusion}\label{sec:conclusion}

In this paper we have introduced an automated approach for detecting relevant patterns in workflow-centric traces with the goal of improving performance debugging in service-based systems.
The approach has been based on a genetic algorithm that aims at clustering traces with similar characteristics, in terms of execution times of called RPCs, that determine latency degradations.
We have applied our approach on a microservice-based application, and we have compared our results (in terms of effectiveness and efficiency) with four other approaches.
The results demonstrate that our approach outperforms existing ones, especially when system runtime behavior is not very regular but varies over time in terms of execution time of called RPCs. This is a relevant result, because non-regular behavior systems are the ones where it is difficult to determine, without automation, the causes of performance degradation. 

Our first promising results encourage us to deeply investigate the application of our approach to other distributed systems, ever more chaotic, so to gain confidence in its ability to capture performance degradation patterns. As future work, by relaying on a wider experimentation, we intend to identify peculiar characteristics of degradation patterns, possibly system-specific, so to proactively provide recommendations to system designers before a performance degradation appears.
We intend to generalize the approach to trace attributes other than RPC execution time (e.g. request size), thus addressing the issue defined in Section \ref{sec:step}.
We also plan to put effort on the improvement of the efficiency and scalability of our approach,
as we believe that precomputed information can be further exploited to reduce the amount of computation needed to identify latency degradation patterns.

\section*{Acknowledgments}
This work was supported by the project "Software Performance in Agile/DevOps context" funded within Programma Operativo Nazionale Ricerca e Innovazione 2014-2020.
\balance
\bibliographystyle{ACM-Reference-Format}
\bibliography{references}

\end{document}

%% file: introduction2.tex
\section{Introduction}
Modern high-tech companies deliver new software in production every day \cite{Feitelson2013} and perceive this capability as a key competitive advantage \cite{Rubin2016}.
In order to support this fast-paced release cycle, IT organizations often employ several independent teams that are responsible "from development to deploy" \cite{OHanlon2006} of loosely coupled independently deployable services.
One challenge of this "move fast" mentality is to ensure software quality \cite{Rubin2016}.
State-of-art performance assurance practices usually involve load testing \cite{Jiang2015},
which unfortunately only covers quite specific use cases,
while performance limits and unexpected behaviors often emerge in the field with live-user traffic \cite{Veeraraghavan2016}.
In this context, performance debugging in production is becoming an essential activity of software maintenance. 
Debugging performance issues can be hard in heterogeneous distributed systems,
where a request may involve several Remote Procedure Calls (RPCs)
and each call can be subject to performance fluctuation due to several reasons (e.g. computational expensive code paths, cache misses, synchronous I/O, slow database queries).
Indeed, since machine-centric tracing mechanisms are often insufficient in these cases,
recent research has developed workflow-centric tracing techniques \cite{Sambasivan2016, Kaldor2017, Sigelman2010}.
These latter techniques capture the workflow of causally related events (e.g., work done to process a request) among the services of a distributed system,
as well as their performance metrics and traced information, e.g., RPCs execution times, HTTP headers, resource consumption, execution nodes or application logs.

Despite the effort in developing techniques to collect causally related performance data,
there is still lack of research on how to exploit workflow-centric traces to provide useful suggestions during performance debugging.
Widely popular workflow-centric technologies (e.g. Zipkin\footnote{https://zipkin.io} , Jaeger\footnote{https://www.jaegertracing.io} , Dapper\cite{Sigelman2010}, etc) provide Gantt charts to debug performance issues related to RPC degradation (see Figure \ref{fig:req}).
Gantt charts are used to show individual requests,
where Y-axis shows the overall request and resulting RPCs issued by the distributed system, and the X-axis shows relative time.
The relative start time and execution time of RPCs are encoded by horizontal bars.
However, Gantt charts effectiveness is restricted to cases where the targets of the analysis are one or few requests.
Indeed, as the number of requests under analysis grows, the causes of performance degradation potentially increase.
For example, the workload of a system can change multiple times during a single day, thus causing performance degradations on different RPCs.
Furthermore, different request parameters can trigger performance deviations on different RPCs.
Specific patterns can \emph{emerge} in RPCs execution times from these complex non-deterministic behaviors,
which we call \emph{latency degradation patterns} in this paper.
They usually concern a small subset of RPCs in which execution time deviation is correlated with request latency degradation.
A rough example of latency degradation pattern is "homepage requests slow down when \verb|getProfile| execution time is over $x$ jointly to \verb|getCart| execution time over $y$".
Each pattern identifies a cluster of requests subject to performance degradation,
that can show similar RPC execution time behavior.
Such clusters enable performance analysts to narrow the scope of the analysis to requests and RPCs that are strictly related to the same performance degradation.
Obviously not all patterns have same relevance, in that
relevant patterns are the ones more frequently occurring upon slow requests and rarely or never present upon fast requests.
In this work, we use a quality metric based on F-score to quantify the relevance of latency degradation patterns.

Unfortunately, the manual identification of highly relevant patterns is often unfeasible due to the large number of requests and RPCs involved. Hence, in this paper we introduce automation in searching those patterns, thus to support performance analyst capabilities during performance debugging.
Our approach models the problem of identifying high quality patterns as an optimization problem.
The problem is solved through the combination of an existing dynamic programming algorithm and a novel genetic algorithm.
The source code of the approach is made available on a public repository \cite{traini2019}.

The proposed technique can be applied to any service-based system that employs a common workflow-centric tracing solution.
We evaluate our approach on a microservice-based application and we show that it can effectively and efficiently detect clusters of requests affected by same artificially injected degradations.
We also compare our approach against a state-of-the-art approach for latency profile analysis \cite{Krushevskaja2013} and general-purpose machine learning clustering algorithms (i.e., kmeans, hierarchical and mean shift).

The paper is organized as follows.
Section \ref{sec:problem} describes the problem that we aim to solve, in both informal and formal ways.
Section \ref{sec:approach} describes our approach and its main components:
dynamic programming algorithm, genetic algorithm and employed optimizations.
The research questions, experimental method, validity and results are described in Section \ref{sec:evaluation}.
Section \ref{sec:step} shows how the described problem can be generalized to deal with traced performance information other than RPC execution time.
Section \ref{sec:relatedwork} reports on related work, while final remarks are presented in Section \ref{sec:conclusion}.

%% file: problem.tex
\begin{figure}
  \includegraphics[width=\linewidth]{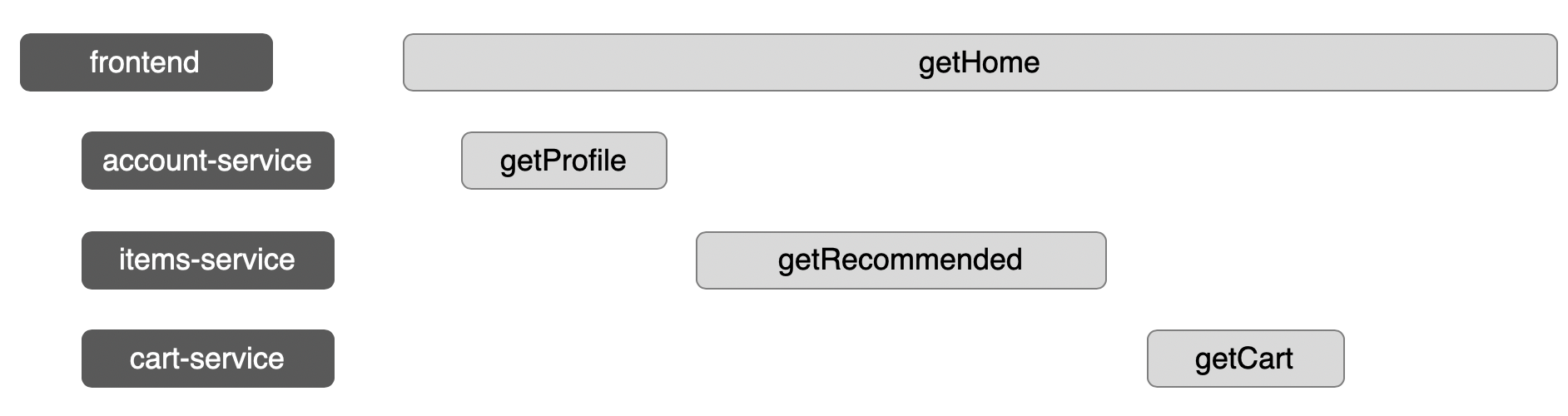}
  \caption{Gantt chart example}
  \label{fig:req}

\end{figure}

\begin{figure}
  \includegraphics[width=\linewidth]{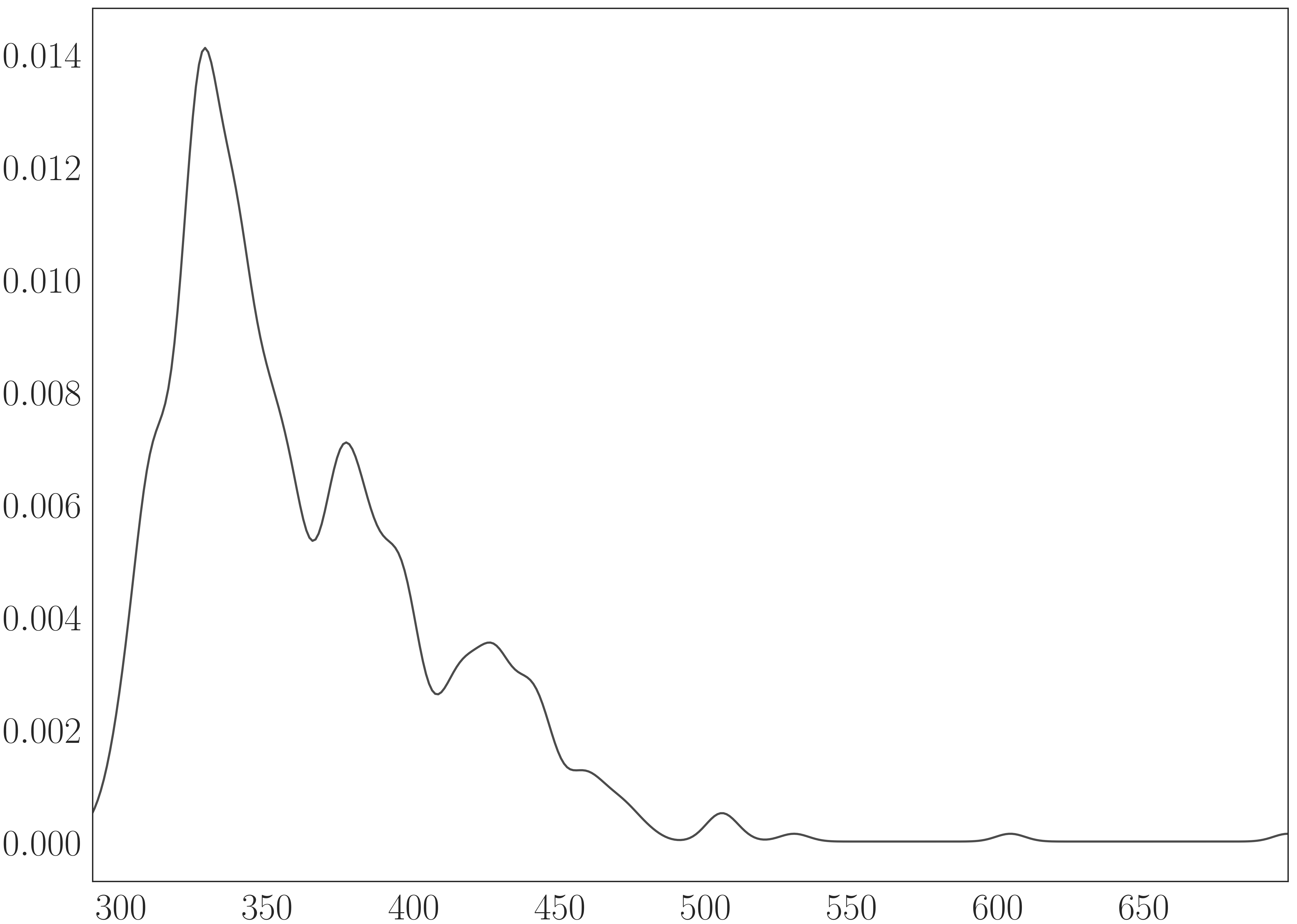}
  \caption{Example of latency distribution.}
  \label{fig:latency}
\end{figure}

\begin{table}[h!]
\centering
	
 \begin{tabular}{|c|c|c|c|c|c|c|} 
 \hline
 \shortstack{Request\\ size\\(bytes)} &
 \shortstack{getRecommended \\ execution time\\(ms)} &
 \shortstack{getCart \\ execution time\\(ms)} & ... &
 \shortstack{getHome\\ latency\\(ms)}\\ [0.5ex]
 \hline\hline
200 & 60 & 60  & ... & 320\\
 \hline
280 & 75 & 90  & ... & 450\\
 \hline
... & ... & ... & ...  & ... \\
 \hline
220 & 60 & 60  & ... & 390\\ [1ex] 
 \hline
\end{tabular}
\caption{Tabular representation of traces}
\label{table:traces}
\end{table}

\section{Problem}\label{sec:problem}
In this section we first informally explain the problem of detecting latency degradation patterns, and 
then we provide a more formal description as an optimization problem.

\input{informalproblem}

\input{formalproblem}

%% file: informalproblem.tex
\subsection{Detecting latency degradation patterns}
Figure \ref{fig:req} shows the Gantt chart example of a homepage request to a service-based e-commerce system.
The workflow starts with the call of \verb|getHome| API exposed by the \verb|frontend| service,
and it involves three RPCs (i.e., \verb|getProfile|, \verb|getRecommended| and \verb|getCart|)
exposed by three different services (i.e., \verb|account|\-\verb|service|, \verb|items-service| and \verb|cart-service|, respectively).\\
Suppose that a performance analyst wants to detect potential causes for latency degradations in homepage requests.
A workflow-centric tracing solution collects information related to \verb|getHome| requests.
This information can be processed and transformed in a tabular format,
where a row represents a request execution and a column represents a particular traced information (see Table \ref{table:traces}).
In this paper, we refer to a row of the table as a \emph{trace} and to a column as a \emph{trace attribute}.
In particular, we will focus on two kinds of attributes, that are the execution time of each RPC (i.e., all central columns of Table \ref{table:traces}), and the latency of the whole request (i.e., the rightmost column in Table \ref{table:traces}).

Figure \ref{fig:latency} shows an example of the \verb|getHome| request latency distribution, as estimated from traces processing.
The estimated distribution shows that several \verb|getHome| requests latencies deviate from the major mode that is around 300-350 milliseconds, hence they represent latency degradations.
Now, let suppose that these degradations are caused by requests that trigger expensive computations in \verb|getRecommended| and a slower database query in \verb|getProfile|.
In this case, a \emph{latency degradation pattern} example could be described as follows:
"\verb|getHome| latency is between 350-670ms when the response time of \verb|getRecommendedItems| is between 70ms and 120ms and the response time of \verb|getCart| is between 80ms and 102ms".
In order to identify such a pattern,
a performance analyst must inspect attributes in all traces, while
comparing requests with latencies that fall within the 350-670ms interval against the ones that fall outside it.
This task can be trivial when the number of traces and attributes are small,
but it becomes overwhelmingly complex as the scale of problem grows.\\
In order to provide the problem intuition,
the example above has been left intentionally simple,
since it associate just one pattern to latency degradation.
However in real-world distributed systems latency degradations are often associated to multiple causes.
For example, Figure \ref{fig:latency} shows two minor modes in latency distribution, 
a single pattern may not identify these two different performance behaviors.
Obviously, this context further complicates the pattern identification.

%% file: formalproblem.tex
\subsection{Problem definition}\label{sec:formaldef}
We first define a set of notations, we then formally describe the optimization problem.
\begin{definition}
A trace $r$, as observed on a specific request, 
is an ordered sequence of attributes $r=(e_0, e_1, ..., e_m, L)$,
where $e_j$ denotes the execution time of an RPC $j$
invoked by the request, and $L$ is the whole observed latency of the request.
\end{definition}

\begin{definition}
A condition $c$ is a triple $c=\langle j, e_{min}, e_{max} \rangle$,
where $j$ is an RPC, and $[e_{min}, e_{max})$ represent an execution time interval for the RPC $j$.
A request $r=(..., e_j, ...)$ satisfies $c$, denoted as $r\vartriangleleft  c$ , if:
\begin{equation*}
 e_{min}\leq e_j< e_{max}
 \end{equation*}

\end{definition}

\begin{definition}\label{def:pattern}
A pattern $P=\{c_0, c_1, ..., c_k\}$ is a set of conditions.
We say that a request $r$ satisfies a pattern $P$, denoted as $r\vartriangleleft p$,
if:
\begin{equation*}
	\begin{aligned}
 		&\forall c\in P, & r\vartriangleleft c
 	\end{aligned}
\end{equation*}
We say that a request $r$ doesn't satisfy a pattern $P$, denoted as $r\ntriangleleft p$,
if:
\begin{equation*}
	\begin{aligned}
 		&\exists c\in P, & r\ntriangleleft c
  	\end{aligned}
\end{equation*}

\end{definition}

Given a set of traces $R=\{r_0, r_1, ..., r_n\}$ and an interval $I$,
where $I$ represents the request latency interval considered as "degraded",
we want to determine the relevance of a pattern $P$ with respect to $I$.
The most relevant pattern would be one that is always satisfied by "degraded traces" 
(i.e., traces whose request latencies fall within $I$) and never satisfied by the other ones. 
More formally, is a pattern $P$ such that:
\begin{equation}
	\begin{aligned}
	 &  r\vartriangleleft P \iff L\in I\\
	\end{aligned}
\end{equation}

Although it is rare to find a pattern $P$ that satisfies the above described property,
it is obvious that some patterns can be more relevant with respect to others.
In order to quantify the relevance of a pattern $P$ with respect to an interval $I$, 
we use an information retrieval quality metric.
We partition traces in two sets, namely positive and negative traces, as follows:
\begin{equation}
	\begin{aligned}
		& R_{pos}=\{r\in R \mid L \in I \}\\
		& R_{neg}=\{r\in R \mid L \not\in I \}\\
	\end{aligned}
\end{equation}

A pattern $P$ identifies true positives and false positives:

\begin{equation}
	\begin{aligned}
		& tp=\{r\in R_{pos} \mid r\vartriangleleft P \} \\
		& fp=\{r\in R_{neg} \mid r\vartriangleleft P \}\\
	\end{aligned}
\end{equation}

Hence, $P$ has precision and recall defined as follows:
\begin{equation}\label{eq:precrec}
	\begin{aligned}
		precision= &\frac{\mid tp \mid}{\mid tp \mid + \mid fp \mid}&\\\\
		recall= & \frac{\mid tp \mid}{\mid R_{pos} \mid }&		
	\end{aligned}
\end{equation}

If a pattern shows high recall, then it means that it frequently appears in positives, although
this does not guarantee correlation with latency degradation,
because identified attribute values can also be present in negative traces.
On the other hand, a high value of precision indicates that, when a trace satisfies the pattern,
its latency usually falls in the latency degradation interval.
However, this does not exclude that traces satisfying the pattern are too few to be worth to investigate.\\
In order to overcome both limitations, as in \cite{Krushevskaja2013},
we use the harmonic mean of precision and recall (i.e., F-score)
to provide a quality score for \emph{latency degradation pattern}:

\begin{equation}\label{eq:fscore}
	Q(P, I)=2\frac{precision\cdot recall}{precision+recall}
\end{equation}

The goal of our optimization problem is to identify a pattern $P^*$ which maximizes the quality score:
\begin{equation}\label{eq:obj1}
	P^*= \arg \max_{P}Q(P, I)
\end{equation}

However, an entire interval $I$ is not likely to be explained by the same pattern,
whereas multiple different latency degradation patterns can be found in traces that 
affect different sub-intervals of $I$.
For this reason, we use the approach proposed in \cite{Krushevskaja2013},
where a set of potential split points $\{s_0, s_1, ..., s_k\}$ is pre-defined 
on an interval $I=[s_0,s_k]$.
This set can be chosen using local minima in latency distribution,
the key insight is to partition the set of request with latencies $L\in I$ in groups that show similar latency behavior.
Furthermore, high density regions of the latency interval (modes) are often associated to the same performance degradation causes\footnote{http://www.brendangregg.com/FrequencyTrails/modes.html}.
For example, in Figure \ref{fig:latency} the second mode (350-400ms) may be related to a cause and the third mode (400-500ms) to another. 

Let $\Theta(P)= \max_{P}Q(P, (s_i, s_j))$ be the score of a sub-interval $(s_i, s_j)$.
The ultimate goal of our approach is to identify a subset of split points  $\{s_0^*, s_1^*, ..., s_z^*\}$ (where $z\leq k$, $s_0^*\equiv s_0$ and $s_z^*\equiv s_k$) that maximize the following equation:

\begin{equation}\label{eq:obj2}
	\sum_{i=0}^{z-1} \Theta(s_i^*, s_{i+1}^*)
\end{equation}

The key intuition is that by optimizing the sum of scores,
the interval $I$ is partitioned in sub-intervals in a way that favors the identification of more relevant patterns over the others.

%% file: approach.tex
\section{Approach}\label{sec:approach}
The main problem of maximizing equation (\ref{eq:obj2}) requires the solution of the sub-problem described by the equation (\ref{eq:obj1}).
In order to solve the main problem we use the dynamic programming approach proposed in \cite{Krushevskaja2013}.
Let $D(i)$ denote the best score for a solution that covers interval $[s_0, s_i)$, with the initial score D(0) = 0. The update step is:
\begin{equation}
	D(i)= D(j) + \max_{0\leq j < i}(\Theta(s_j, s_i) ) 
\end{equation}
Hence, to construct the solution that covers interval $[s_0, s_i)$ the algorithm search for each possible pair $i$, $j$ such that $D(j) + \Theta(s_j, s_i)$ is maximized.

In the following we describe how we solve the sub-problem, which represents the core novelty of our approach.
We first describe the components of our search-based approach,
and then we describe how fitness evaluation is optimized through search space reduction and precomputation.

\subsection{Genetic algorithm}
Our approach uses Search Based Software Engineering (SBSE) \cite{Harman2012}, an approach in which software engineering problems are reformulated as search problems within the search space that can be explored using computational search algorithms.
Specifically, we use a Genetic Algorithm (GA).
GAs are based on the mechanism of the natural selection \cite{Holland1992} and they use stochastic search techniques to generate solutions to optimization problems. The advantage of GA is in having multiple individuals evolve in parallel to explore a large search space of possible solutions.

Our GA is implemented on top of the DEAP framework \cite{Fortin2012}.
In the following we describe the five key ingredients of our GA implementation (i.e., representation, mutation, crossover, fitness function and computational search algorithm) in the context of our sub-problem defined in equation (\ref{eq:obj1}).

\textbf{Representation}: Feasible solutions to the sub-problem are all possible patterns $P$.
We recall that a pattern is defined as a set of conditions $P=\{c_1, c_2, ..., c_k \}$, where each condition is a triple $<j, e_{min}, e_{max}>$, with $j$ referring to the RPC subject to the condition and $[e_{min}, e_{max})$ representing the execution time interval.

\textbf{Mutation}: The mutation randomly choose among three mutation actions, namely: add, remove or modify.
The first action adds a new randomly generated condition to the pattern $P$. However, if the RPC involved in the new condition is already present in another condition contained in $P$ then the mutation doesn't have any effect.
The remove mutation randomly removes a condition from $P$.
The modify mutation randomly selects a condition $c$ and modifies one of the two endpoints of the interval,
and then reorders them, if necessary, so that $e_{min}<e_{max}$.

\textbf{Crossover}: Given two patterns $P_1$ and $P_2$, the two sets of condition are joined together $P_U = P_1\cup  P_2$ and then randomly partitioned in two new patterns $P'_1$ and $P'_2$.

\textbf{Fitness function}:
In order to evaluate the fitness of each solution we adopt the quality score described by equation (\ref{eq:fscore}).
However, the computation of such score can be overwhelmingly expensive,
hence we optimize the fitness evaluation with the techniques described in next subsection.

\textbf{Computational search}:
We use a ($\mu + \lambda$) genetic algorithm \cite{Beyer2002}.
As a selection operator, we use a tournament selector \cite{Koza1992} with tournament size by 20.
Crossover and mutation rates are fixed to 0.8 and 0.2,
whereas $\mu$ and $\lambda$ are both set to 100.
The evolutionary process terminates after 400 generations with a 100 population size.

\subsection{Optimization of fitness evaluation}\label{sec:fitness-optimization}
Fitness evaluation is one of the most frequently executed operations during the evolutionary process.
A time-consuming fitness evaluation can severally hamper the approach efficiency.
The presented approach uses precomputation to enhance fitness evaluation performance, where
the employed technique requires a search space reduction.
Although search space reduction can potentially cut off optimal or near-optimal solutions,
we employ a smart reduction policy which still preserves search space quality.
In the following we explain the key idea behind our precomputation technique,
then we illustrate the search space reduction policy.

\begin{figure}
  \includegraphics[width=\linewidth]{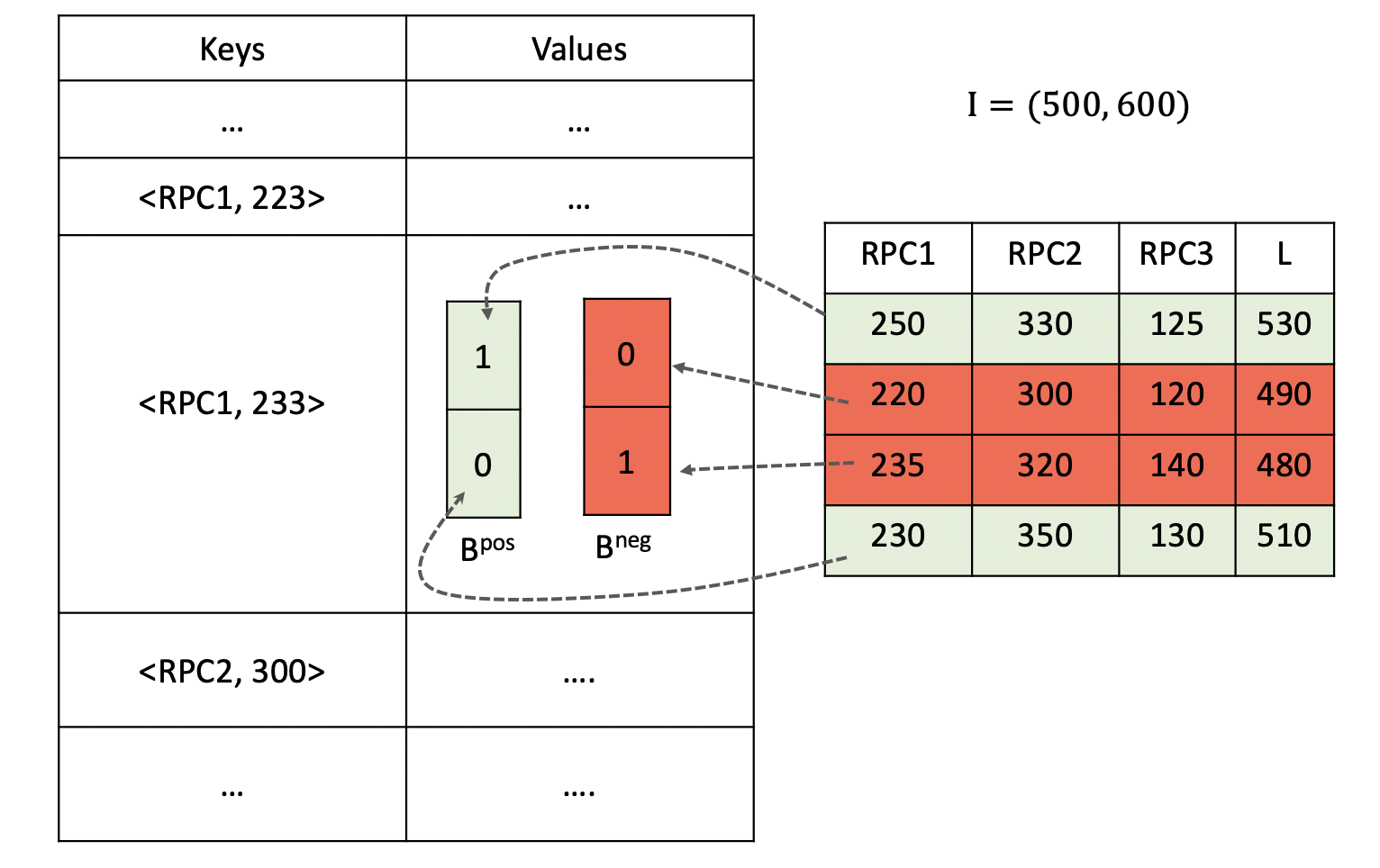}
  \caption{Hash table entry for inequality check <RPC1, 233> over a set of traces and an interval I=(500, 600)}
  \label{fig:hashtable}
\end{figure}

\textbf{Precomputation}:
The identification of true positives and false positives for a given pattern $P$ is the most performance-critical operation executed during fitness evaluation.
This operation requires to verify for each $r\in R$ if $r\vartriangleleft P$.
The verification of this property involves a bunch of \emph{inequality checks},
which are likely to be repeated several times during the evolution process.
The aim of our technique is to reduce fitness evaluation effort
by precomputing \emph{inequality check} results in order to avoid redundant computations.
We denote inequality checks as pairs $\langle j, e_t\rangle $, where $j$ is a RPC and $e_t$ is an execution time threshold.
Inequality check results are represented as ordered sequences of booleans $B=\langle b_0, b_1, ..., b_n\rangle $,
where $b_i$ refers to the check result for the trace $r_i\in R$.
A check result $b_i$ for a given inequality check $\langle j, e_t\rangle $ and a trace $r_i= (..., e_j, ...)$ is defined as:
\begin{equation*}
	b_i=\begin{cases}
			True & \text{if $ e_j \geq e_t $}\\
    		False & \text{otherwise}
  		\end{cases}
\end{equation*}
Since the aim is to both improve true positives and false positives computation,
inequality check results are precomputed on positive and negative traces  (i.e., $R_{pos}$ and $R_{neg}$).
Hence, for each inequality check $\langle j, e_t\rangle $ two boolean sequences are generated (namely $B^{pos}$ and $B^{neg}$),
which represent inequality check results, respectively, for positive and negative traces.
Boolean sequences are encoded as bit strings and stored in a hash table,
where the key is an inequality check $\langle j, e_t\rangle $ and the value is a pair of bit strings $\langle B^{pos}, B^{neg}\rangle$.
Figure \ref{fig:hashtable} shows a simplified example of a hash table entry for an inequality check over a set of traces and interval $I$.

These data structures enable fast identification of true positives and false positives across multiple traces through bitwise operations.
A bitwise operation works on one or more bit strings at the level of their individual bits.
We use two common bitwise operators: \emph{and} and \emph{not}.
A bitwise \emph{and} ($\wedge$) is a binary operation that takes two equal-length bit strings and performs the logical \emph{and} operation on each pair of the corresponding bits:
\begin{equation*}
	X \wedge Y = \langle x_1\wedge y_1, x_2\wedge y_2, ..., x_n\wedge y_n\rangle
\end{equation*}
A bitwise \emph{not} ($\neg$) is a unary operation that performs logical negation on each bit:
\begin{equation*}
	\neg B =  \langle \neg b_1, \neg b_2, ..., \neg b_n \rangle 
\end{equation*}
A condition $c=\langle j, e_{min} , e_{max} \rangle$,
can be efficiently evaluated on positive traces as well as on negative traces with the following two steps.
First, boolean sequences associated to inequality checks $\langle j, e_{min} \rangle$ and  $\langle j, e_{max} \rangle$ are retrieved from the hash table.
We denote them as $\langle B_{min}^{pos} , B_{min}^{neg}\rangle$
and $\langle B_{max}^{pos} , B_{max }^{neg}\rangle$ respectively.
Then, positive and negative traces satisfying condition $c$ are derived through bitwise operations:
\begin{equation*}
	B^{pos}_{c} = B_{min}^{pos} \wedge \neg B_{max}^{pos}
\end{equation*}
\begin{equation*}
	B^{neg}_{c} = B_{min}^{neg} \wedge \neg B_{max}^{neg}
\end{equation*}
Where $b_i\in B^{pos}_{c}$ (resp. $b_i\in B^{neg}_{c}$) denotes if $r_i\vartriangleleft c$
with $r_i\in R_{pos}$ (resp. $r_i\in R_{neg}$).\\
The same approach is also applied for pattern satisfaction, $r_i\vartriangleleft P$:
\begin{equation*}
	\begin{aligned}
		& B^{pos}_{P}=\bigwedge\limits_{c\in P} B_c^{pos} \\
		& B^{neg}_{P}=\bigwedge\limits_{c\in P}B_c^{neg} \\
	\end{aligned}
\end{equation*}
Number of true positives and false positives are then obtained by counting $True$ booleans (i.e. number of 1 in the bit string) in both $B^{pos}_{P}$ and $B^{neg}_{P}$:
\begin{equation*}
	\begin{aligned}
		& |tp|= |\{b\in B^{pos}_{P} \mid b=True\}| \\
		& |fp|= |\{b\in B^{neg}_{P} \mid b=True\}| \\
	\end{aligned}
\end{equation*}
Finally, fitness  is derived through a simple numerical computation (see equations (\ref{eq:precrec}) and (\ref{eq:fscore})).

\textbf{Search space reduction}: Since the execution time is a continuous value,
there is an uncountable number of possible inequality checks $\langle j, e_t \rangle$.
Hence, precomputing results for any possible inequality check $\langle j, e_t\rangle$ is unfeasible.
For this goal, we employ a search space reduction to decrease precomputation effort as well as the amount of inequality check results (i.e. bit strings) to store.
Obviously, search space reduction can be risky,
since optimal or near-optimal solutions can be excluded by the search.
We tackle this problem by selecting, for each RPC, only meaningful thresholds $e_t$ according to execution time distribution.
We select thresholds that separate high density regions of the execution time interval.
Those values identify relevant points that should cluster together requests with similar execution time behavior in a certain RPC.
The key intuition is that if execution time of a certain RPC interval is correlated with a relevant latency request degradation,
then its behavior must be recurrent to a relevant number of requests (hence, to a dense interval of the execution time).
Furthermore, RPC execution time distribution often shows multimodal behavior and modes can be often related to performance degradation (e.g., cache hit/miss, slow/fast queries, synchronous/asynchronous I/O, expensive code paths).
Our approach employs a mean shift algorithm \cite{Comaniciu2002} to identify high density intervals of RPC execution time.
Mean shift is a non-parametric feature-space analysis technique for locating the maxima of a density function \cite{Cheng1995},
and its application domains include cluster analysis in computer vision and image processing \cite{Comaniciu2002}.
For each RPC, we cluster traces with the mean shift algorithm according to the corresponding execution time,
we then infer thresholds according to identified highly dense regions.

%% file: evaluation.tex
\begin{figure}
  \includegraphics[width=\linewidth]{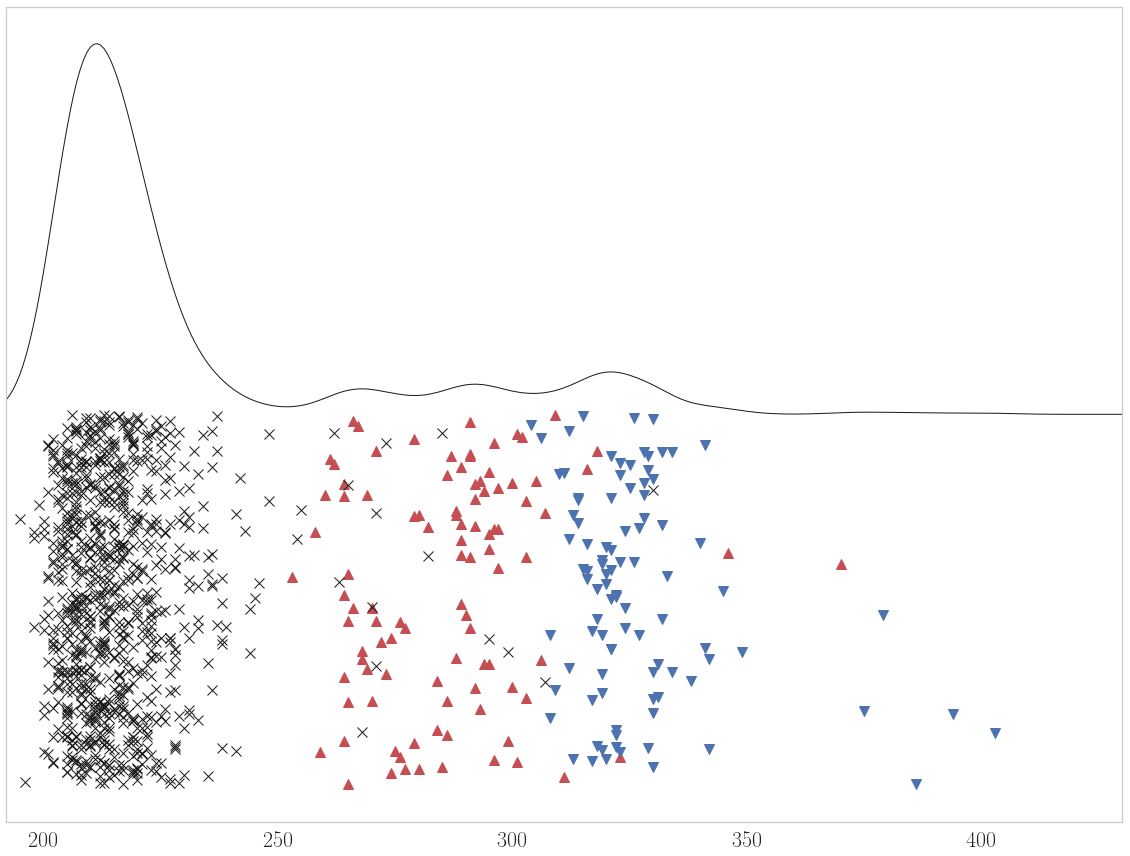}
  \includegraphics[width=\linewidth]{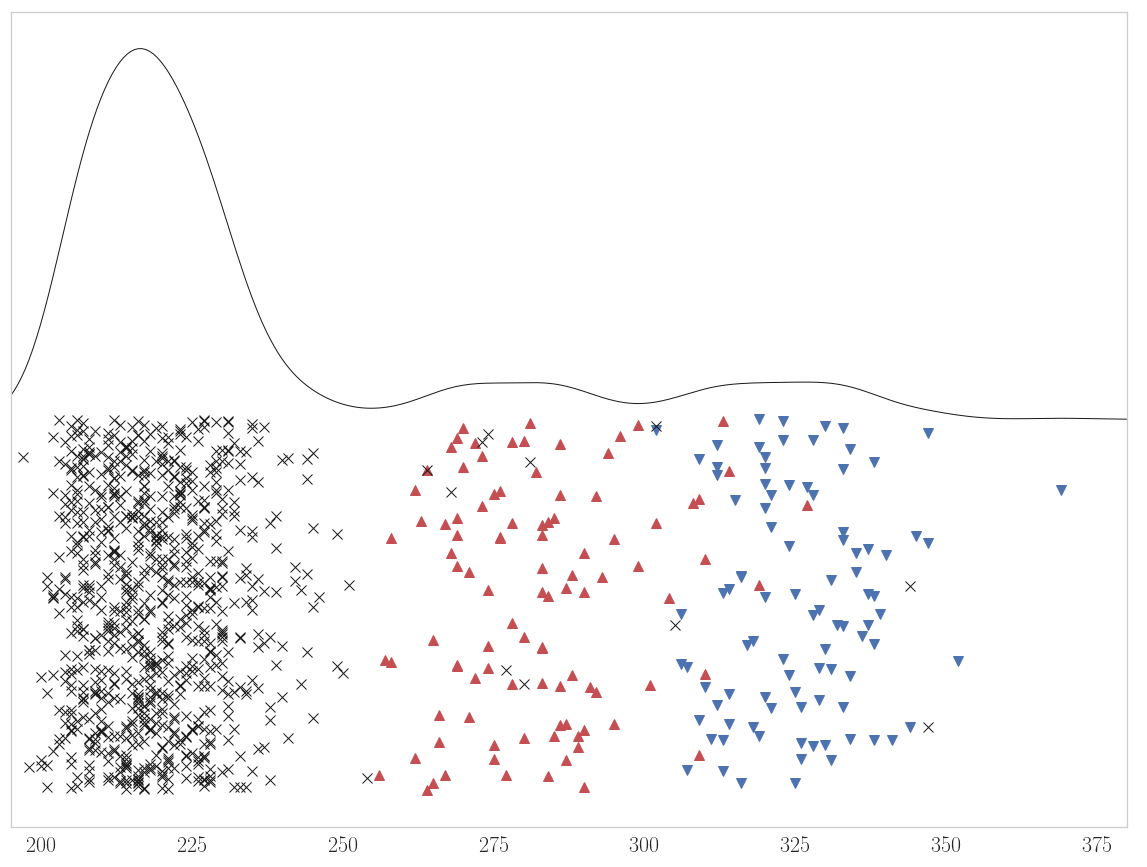}
  \caption{Estimated latency distribution of requests and nominal request latencies}
  \label{fig:experiment}
\end{figure}

\section{Evaluation}\label{sec:evaluation}
In this section, we state our research questions (RQs) and we present the evaluation of our approach on a microservice-based application case study.
We chosen microservices because they represent a widely used paradigm in nowadays service-based systems.
In addition we envision that our approach can be extremely useful to debug performance issues in microservice-based applications,
given high frequency of deployments and continuous experimentation \cite{Schermann2018}(e.g., canary and blue/green deployment).

\subsection{Research questions}

We aim at addressing the following research questions:

\textbf{RQ1} Is our approach effective for clustering requests associated to the same latency degradation pattern, as compared to machine learning algorithms?\\
In order to answer this question, we compare our approach against three general-purpose machine learning clustering algorithms (i.e. K-means, Hierarchical, Mean Shift) that are described in Section \ref{sec: benchmarks}. The rationale beyond this question is the widespread of modern machine learning tools and libraries for clustering problems.

\textbf{RQ2} Is our approach effective with respect to state-of-the-art approaches for latency profile analysis?\\
With this respect, we have identified the work in \cite{Krushevskaja2013} as the closest one to our approach, also described in Section \ref{sec: benchmarks}. The differences are that they adopt a branch and bound algorithm and they target a more general problem, because the attributes that they consider are not limited to latencies. We have implemented their approach for sake of result comparison.

\textbf{RQ3} How robust is our approach to "noise"?\\
We have introduced two types of \emph{noise} in our experiments, which are described in Section \ref{sec:Method} and can affect detection capabilities of our approach. We have compared the above mentioned approaches in terms of their effectiveness in presence of noise.

\textbf{RQ4} What is the efficiency of our approach as compared to other ones?\\
This question strictly concerns the execution time. We have measured the execution time of all considered approaches applied on the case study, for sake of a costs/benefits analysis.

\begin{figure}
  \includegraphics[width=\linewidth]{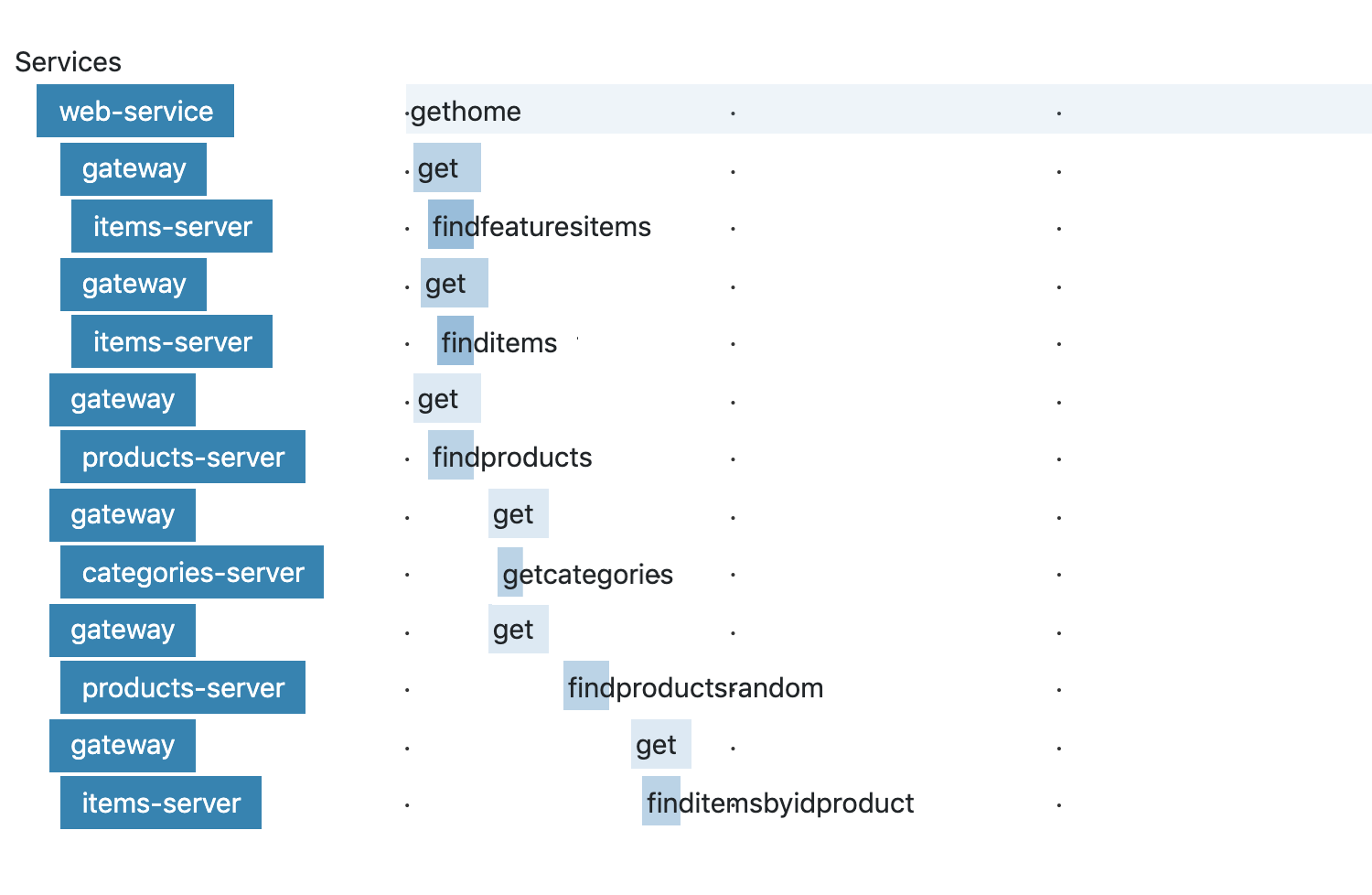}
  \caption{Gantt chart of a request loading the homepage of E-Shopper}
  \label{fig:zipkin}
\end{figure}

\subsection{Subject application}
We experiment our approach on E-Shopper\footnote{https://github.com/SEALABQualityGroup/E-Shopper},
that is an e-commerce micro\-services-based web application.
The application is developed as a suite of small services, each running in its own Docker\footnote{https://www.docker.com} container and communicating via RESTful HTTP APIs.
It is composed by 9 microservices developed on top of the Spring Cloud\footnote{https://spring.io/projects/spring-cloud} framework,
where each microservice has its own MariaDB\footnote{https://mariadb.org/} database.
The application produces traced data that are reported and collected by Zipkin\footnote{https://zipkin.io}, i.e. a popular workflow-centric tracing system,
and stored in Elasticsearch\footnote{https://www.elastic.co}.
Traced information are then processed and transformed in a tabular format, such as the one showed in Table \ref{table:traces}.
Each trace refers to a single request, while trace attributes are RPC pure execution time.
By pure execution time we mean the RPC execution time minus the time waiting for invoked RPCs to terminate.
For example, in Figure \ref{fig:req} the RPC \verb|getHome| calls three synchronous RPCs (\verb|getProfile|,\verb|getRecommended|, \verb|getCart|),
hence pure execution time of getHome is its execution time minus the time waiting for termination of the three invoked RPCs.
Note that asynchronous calls do not introduce waiting time, thus we do not consider them in pure execution time calculation.
We focus our analysis only on requests loading the homepage,
since it is the request which trigger more RPCs.
Specifically, each request involves 13 calls of 8 unique RPCs among 5 microservices,
as showed by the Gantt chart showed in Figure \ref{fig:zipkin}.
Note that the first two \verb|get| RPCs  invoked by \verb|web-service| are asynchronous,
hence they do not block \verb|gethome| execution.

\subsection{Methodology}
\label{sec:Method}
The main goal of our empirical study is to determine whether the presented approach is able to identify clusters of requests affected by the same degradation causes.
In order to achieve this goal, we perform multiple load test sessions in which we inject recurrent \emph{artificial degradations}, thereafter we run our approach on each set of collected traces to evaluate whether clusters of requests affected by same artificial degradations are correctly identified.
In our empirical study, artificial degradations are actualized as delays injected in RPCs.
Before each session, we randomly define two recurrent artificial degradations $A_1$ and $A_2$,
for example:
\begin{itemize}
	\item $A_1$: 50ms delay added to \verb|findfeaturesitems| RPC and 50 ms delay added to \verb|getcategory| RPC;
	\item $A_2$: 50ms delay added to \verb|gethome| RPC. 
\end{itemize}

Then, during the load test, requests are randomly marked as \emph{affected} by one of the two artificial degradations with 0.1 probability.
When a request is marked as affected,
the mark is propagated through RPCs, thus leveraging context propagation \cite{Mace2018},
and delays are injected in RPCs according to the artificial degradation definition.
Hence, requests affected by the same artificial degradation will always have the same delays injected on the same RPCs.
At the end of each load test session, each artificial degradation approximately affects 10\% of requests.
Figure \ref{fig:experiment} shows the latency behavior of the main RPC \verb|gethome| (i.e., the target of our analysis)
during a load test session.
Specifically, it shows the shape of the latency distribution and, under the curve, the latencies of nominal requests randomly distributed on the Y-axis. Time is expressed on X-axis in milliseconds. 
Requests not affected by any artificial degradation are represented as black x,
the ones affected by the artificial degradation $A_1$ are represented as blue down triangles,
whereas red up triangles represent requests affected by $A_2$.

Each load test session lasts 5 minutes and involves a synthetic workload simulated by Locust\footnote{https://locust.io/},
which makes a request to the homepage every 50 milliseconds.
Since not all RPCs are synchronous, some injected delays may not cause latency degradation.
In order to avoid these cases, only synchronous RPCs are considered in artificial degradations.
In our experiments we consider three types of artificial degradations:
\emph{type 1} injects a delay to just one RPC,
\emph{type 2} injects delay to two RPCs and \emph{type 3} to three RPCs.
Each injected delay slows down RPC execution time by 50ms.
Artificial degradations by the same type are expected to produce a similar performance degradation in terms of request latency,
since the total amount of injected delays is the same.\\
Before each load test session, the pair of artificial degradation $A_1$ and $A_2$ is generated as follows.
First, the number of RPCs affected by both artificial degradations is defined
by randomly assigning \emph{types} to $A_1$ and $A_2$, respectively.
The assignment is made by ensuring that $A_1$ and $A_2$ always have different types,
i.e. they produce a different performance degradation in terms of request latency (see Section \ref{sec:threats} for more details).
Then, RPCs affected by delays are randomly selected, among the 6 synchronous RPCs (see Figure \ref{fig:zipkin}), for each artificial degradation according to their types.


Distributed systems are often noised,
hence in order to test the robustness of our approach we also performed "noised" load testing sessions.
In particular, we consider two types of noises:
\begin{itemize}
\item The first noise is a small deviation of delays injected in RPCs.
The key insight is to reproduce situations where performance degradation doesn't shows a constant behavior.
For each artificial degradation, a random RPC is chosen among the affected ones so that
delays injected on this RPC will not have a constant behavior,
i.e.  the delay injected is 60ms instead of 50ms in half of the requests.
\item The second type of noise involves situations where RPCs execution time degradation doesn't cause
any latency degradation on the overall request.
In particular, for each artificial degradation we select one of the two asynchronous calls (i.e., \verb|findfeaturesitems| and \verb|finditems|) and inject a 100ms delay in half of requests to this call affected by artificial degradation. We have preliminarly experimented that those delays do not cause slow downs in requests.
\end{itemize}
We have performed 10 different load testing sessions with randomly generated artificial degradations, where 5 sessions are noised.
This has generated 10 different sets of traces.
We then ran our approach as well as baseline approaches to evaluate their effectiveness.
For each load test session, we targeted a specific latency degradation interval $I$.
We have chosen, for each session, the interval $(L_{min}, L_{max})$, where $L_{min}$ and $L_{max}$ are the minimum and the maximum observed latency of \emph{affected} requests in the session.
Our approach, as well as the one proposed by Krushevskaja and Sandler \cite{Krushevskaja2013},
requires as input a set of potential split points $\{ s_0, s_1, ... s_k\}$.
In each experiment, we identify the set of split points by using the same approach used in search space reduction to identify thresholds of RPC execution time (see Section \ref{sec:fitness-optimization}),
i.e. local minima identified through Mean shift algorithm.
For the considered clustering algorithms (i.e., Kmeans, Hierarchical and Mean shift), we use as inputs only traces that fall in the target interval $I$, since the goal is to cluster degraded requests with same artificial degradations. We also set a predefined number of clusters for Kmeans and Hierarchical, we run these algorithms multiples times with different inputs (i.e., k=2,...,6), and we then pick the best achieved solution for each set.

The output of each approach is a set of clusters.
We select, for each injected artificial degradation, the best matching cluster for every approach, 
by identifying best pairs of cluster and artificial degradation $\langle C_i, A_i \rangle$, such that F-score is maximized while considering requests affected by artificial degradation $A_i$ as positives.
At the end of this process, for each approach we have two best clusters $C_1$ and $C_2$,
each one associated to the respective artificial degradation $A_1$ and $A_2$.  

Finally, we then evaluate the effectiveness of each approach by using the following metrics.
In theory, an ideal approach would identify clusters that correspond to the group of requests affected by same artificial degradation (i.e. blue triangles and red triangles in Figure \ref{fig:experiment}).
We evaluate each approach effectiveness in terms of recall, precision and F-score.
An approach with low recall would not be adopted in practice, since it fails to detect relevant latency degradation patterns.
An approach that produces results with high recall and low precision is not useful either,
since identified clusters do not precisely correspond to the same artificial degradation.
The three evaluation metrics are formally defined as follows.
Let us name $G$ the subset of requests that are correctly associated to the corresponding artificial degradation, 
$P$ the subset of requests affected by one of the two artificial degradation, $C_1$ and $C_2$ the two identified clusters.
 We define the recall as $|G|/|P|$, the precision as $|G|/(|C_1|+|C_2|)$, and the F-score as from equation \ref{eq:fscore}.

\subsection{Threats to Validity}\label{sec:threats}
A threat to validity of our empirical study is that our experiments were performed on only one subject application,
which makes it difficult to generalize the results to other distributed service-based systems.
However, E-Shopper has been already used as a representative example of a microservice-based application in software performance research \cite{Arcelli2019}. We expect our results to be generalizable to other distributed systems that employ a common workflow-centric tracing solution.

Our current implementation only consider RPCs execution time for cluster identification,
whereas often other trace attributes are correlated to request latency degradation (such as resource consumptions, http headers, RPCs execution node, etc.).
While this is a potential threat, in our opinion this is not a major one, since our approach can be easily adapted to other attribute types.
There is no theoretical limitation that prevents our approach from considering attributes other than RPCs execution time.
A first formalization in this direction is discussed in Section \ref{sec:step}.

Artificial delays were injected into randomly chosen RPCs at application-level.
This may be a threat,
since performance degradation can happen on different level of the software stack (e.g., libraries, operating systems, databases, networks, etc.), and it can be due to different reasons, such as workload spikes, network issues, contention of hardware resources and so on.
However, in this work we only consider performance degradations related to RPC execution time,
hence we consider injected delays as a rough but reasonable simulation of a generic performance problem (e.g. slow query, expensive computation, etc).
Also, in our experiments we only consider cases where injected pairs of artificial degradations have different types. 
Basically, we do not consider cases where different performance issues produce same performance degradation in terms of request latency.
Those cases are excluded from evaluation, because they are very specific and less frequent in practice.
However, we plan to extend our study also considering these cases in future works.

A different threat is that we perform load test sessions with a single synthetic user.
We used such a simple and controllable workload,
because it allows us to have control on causes of relevant latency degradations.
Using a more intense and mixed workload (e.g. requests to different pages of the application) may lead to more chaotic system behavior,
but injected performance degradation may become not relevant and the approach difficult to evaluate.
We leave the evaluation of our approach in more chaotic contexts to future works.
Also, the approach is evaluated on sets of about 1000 traces,
while performance debugging in modern distributed systems could involve higher number of traces.
We plan to evaluate the scalability of the approach in future.

In spite of these threats, this empirical study design allowed us to evaluate our approach in a controlled setting. Thus, we are confident that the threats have been minimized and our results are reliable.

\subsection{Baseline approaches}\label{sec: benchmarks}
In this section we describe the approaches that we have used as baselines, that are:
three widely popular clustering algorithms (i.e., K-Means, Hierachical, Mean shift) and an optimization based approach (that we call here Branch and Bound \cite{Krushevskaja2013}).
For clustering algorithms we use the implementation provided by scikit-learn
Machine Learning library\footnote{https://scikit-learn.org}.
We have instead implemented the optimization approach,
since no implementation was available, and
the source code is publicly available in \cite{traini2019}.

\textbf{K-Means} The K-Means algorithm \cite{macqueen1967} clusters data by trying to separate samples in $k$ groups by equal variance, while minimizing a criterion known as within-cluster sum-of-squares.
KMeans requires the number of clusters to be specified.

\textbf{Hierachical}
Hierarchical clustering \cite{Rokach2005} is a general family of clustering algorithms that build nested clusters by merging or splitting them successively. We use an implementation based on a bottom up approach: 
each observation starts in its own cluster, and clusters are successively merged together.
Also hierarchical clustering requires the number of clusters to be specified.

\textbf{Mean shift}
MeanShift clustering \cite{Comaniciu2002} aims to discover blobs in a smooth density of samples. It is a centroid based algorithm, which works by updating candidates for centroids to be the mean of the points within a given region.
The mean shift algorithm doesn't require upfront specification of number of clusters.

\textbf{Branch and bound}
The branch and bound approach proposed in \cite{Krushevskaja2013} is the closest one to our work.
It aims to explain latency intervals through combination of traces attributes.
It works both on continuous and categorical trace attributes.
In case of continuous attributes, an encoding step is required to transform them in a binary form:
the value range of an attribute is split between several binary attributes, each one corresponding to an interval of value.
In our experiments, we split the RPC execution time by using the same approach used in search space reduction (i.e. Mean shift algorithm, see Section \ref{sec:fitness-optimization}).

%% file: results.tex
\begin{figure*}
  \includegraphics[width=\linewidth]{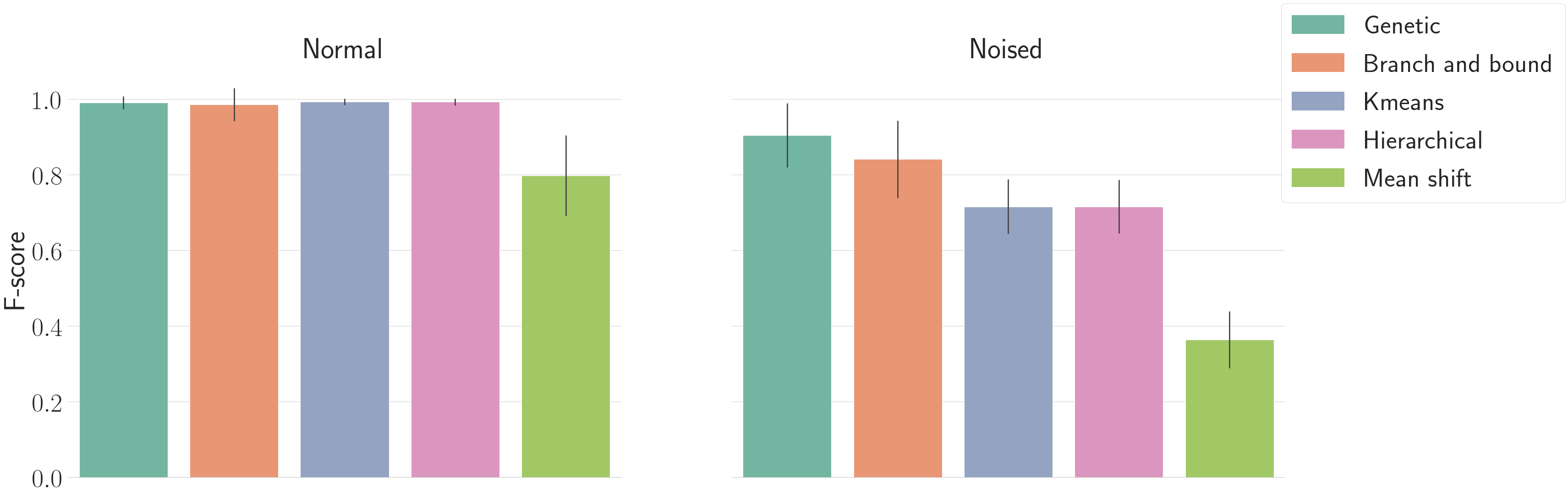}
  \caption{F-score in experiments: average and standard deviation}
  \label{fig:fscore}
\end{figure*}

\begin{figure*}
  \includegraphics[width=\linewidth]{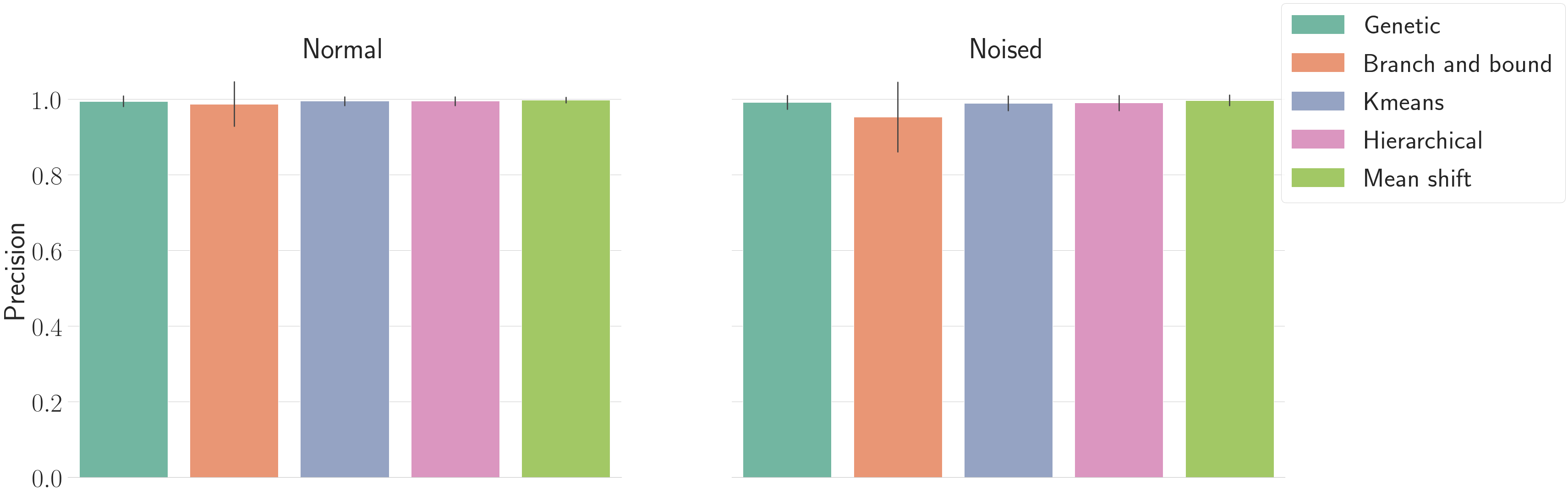}
  \caption{Precision in experiments: average and standard deviation}
  \label{fig:prec}
\end{figure*}

\begin{figure*}
  \includegraphics[width=\linewidth]{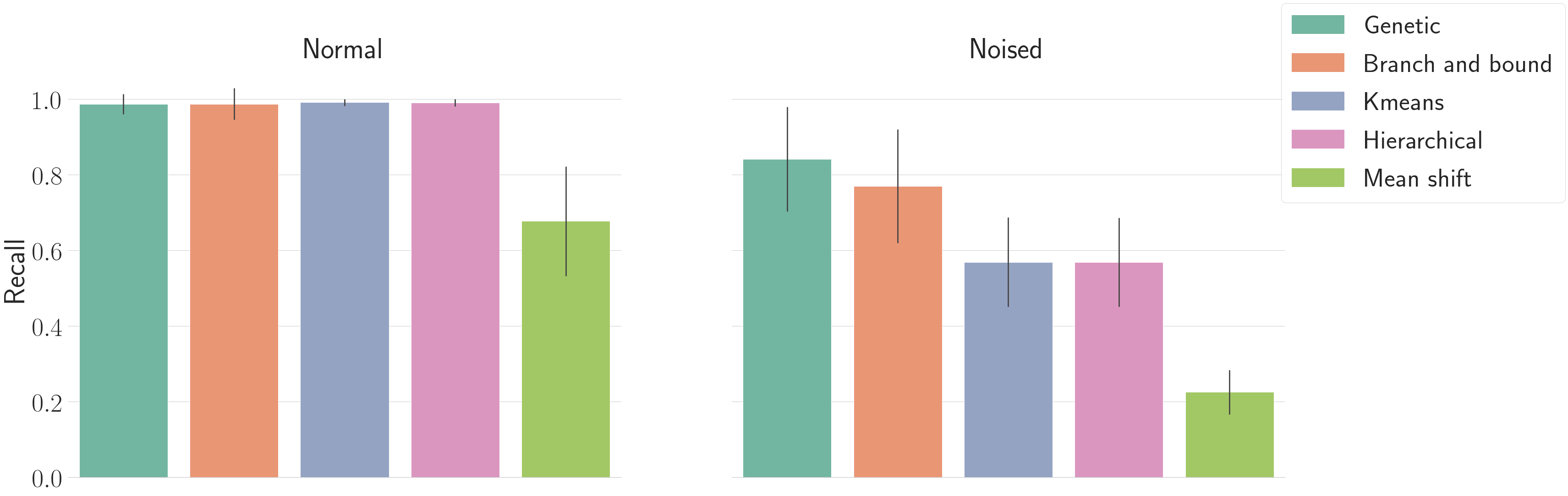}
  \caption{Recall in experiments: average and standard deviation}
  \label{fig:rec}
\end{figure*}

\subsection{Results} 
Figure \ref{fig:fscore} shows average and standard deviation of F-scores achieved by each approach in both normal and noised experiments,
where our approach is denoted as \emph{Genetic}.
F-score in normal experiments suggests that all the approaches identify near optimal clusters, except for the Mean shift algorithm.
Conversely, results of noised experiments show different performances among approaches.
According to reported F-scores in noised experiments (see detailed results in Table \ref{table:results}),
clusters identified by our approach are better than those identified by machine learning clustering approaches. 
Kmeans, Hierarchical and Mean shift are highly precise (see Figure \ref{fig:prec}), 
thus suggesting that each identified cluster is almost always composed by requests affected by the same artificial degradation.
But their performances are low in terms of recall (see Figure \ref{fig:rec}),
in that an average recall of $\approx 0.57$ for Kmeans and Hierarchical,
and of $\approx 0.22$ for the Mean Shift algorithm have been obtained.
These results suggest that identified clusters do not include significant portion of requests affected by artificial degradations.
The reason of these behaviors relies on the fact that machine learning approaches blindly group together requests 
with similar RPC execution times, without considering their correlation with latency degradation.
Therefore, performance fluctuation in RPC that doesn't produce any effect in the request,
such as performance degradations in asynchronous RPC (i.e. second type of noise), can easily confuse those methods.
Our approach provides on average +48\% in terms of recall with respect to best clustering approaches (Kmeans and Hierarchical),
without showing decrease in precision,
thus leading to an improvement +26\% in terms of F-score, thereby positively answering to RQ1.

\input{table}

The analysis of the F-score (see Figure \ref{fig:fscore}) reveals that our approach
not only outperforms common machine learning approaches, but also the state-of-the-art technique against which we compare it.
Our approach shows similar precision with respect to the branch and bound approach (see Figure \ref{fig:prec}),
but it achieves improvements in terms of recall.
The result of Wilcoxon test confirms that our approach outperforms branch and bound in noised experiment in terms of recall with statistical significance ($p < 0.05$) and medium effect size (Cohen's d $< 0.8$ and $\geq 0.5$).
Although both approaches are driven by the same optimization objective,
the presented technique achieves an improved recall due to the searching capability of the genetic algorithm on a wider solution space.
To further confirm the improvement over the state of art,
we also performed Wilcoxon test on F-Score results.
Results of statistical test confirm that our approach outperforms branch and bound approach with statistical significance ($p < 0.05$) in both normal and noised experiments,
where the effect size is negligible ($< 0.2$) in normal experiments and medium in noised experiments.
These results suggest that performance analysts should use our approach for clustering requests affected by similar performance issues,
since identified clusters are more comprehensives than those identified by the state of the art approach,
thus positively answering to RQ2.

Both optimization-based approaches seems to be more resilient to noise with respect to machine learning approaches.
Optimization-based approaches steer the search towards clusters that are strictly related to latency degradation.
Comparison of F-score among normals and noises experiments show the robustness of our approach:
Genetic approach decreases its average F-score of just $\approx 8\%$, Branch and Bound decreases its average F-score of $\approx 14\%$, Hierachical and Kmeans of $\approx 28\%$, and the Meanshift of $\approx 54\%$. 
These results show the robustness of our approach, hence positively answering to RQ3.

\begin{table*}[]
\small
\begin{tabular}{c|c|c|c|c|}
\cline{2-5}
                                       & \multicolumn{2}{c|}{Normal}                                                                                                                      & \multicolumn{2}{c|}{Noised}                                                                                                                         \\ \hline
\multicolumn{1}{|c|}{Approach}         & \begin{tabular}[c]{@{}c@{}}Execution time\\ mean (sec)\end{tabular} & \begin{tabular}[c]{@{}c@{}}Execution time\\ standard deviation (sec)\end{tabular} & \begin{tabular}[c]{@{}c@{}}Execution time\\ mean (sec)\end{tabular} & \begin{tabular}[c]{@{}c@{}}Execution time\\ standard deviation (sec)\end{tabular} \\ \hline
\multicolumn{1}{|c|}{Genetic}          & 17.551                                                              & 3.390                                                                              & 37.106                                                              & 21.143                                                                            \\ \hline
\multicolumn{1}{|c|}{Branch and bound} & 49.070                                                               & 12.820                                                                             & 102.259                                                             & 89.291                                                                            \\ \hline
\multicolumn{1}{|c|}{Kmeans}           & 0.024                                                               & 0.006                                                                             & 0.025                                                               & 0.005                                                                             \\ \hline
\multicolumn{1}{|c|}{Hierachical}      & 0.004                                                               & 0.001                                                                             & 0.004                                                               & 0.001                                                                             \\ \hline
\multicolumn{1}{|c|}{Mean shift}       & 0.509                                                               & 0.08                                                                              & 0.307                                                               & 0.039                                                                             \\ \hline
\end{tabular}
\caption{Execution time in experiments: average and standard deviation}
\label{table:time}
\end{table*}

Table \ref{table:time} shows the average execution time and standard deviation of each approach.
Note that each load test session produces a set of about 1000 traces,
therefore the execution time results are referred to this scale.
Common clustering algorithms are extremely faster compared to optimization-based methods.
However they fail to achieve a good recall in noised experiments.
Furthermore, both optimization-based methods provide as output, together with each cluster, a description of its main characteristics (i.e. latency degradation pattern),
which can then be used as a valuable starting point for a deeper performance analysis.
Performance analyst should decide, based on the context, which technique to use.
Machine learning clustering are preferable only
in cases where optimizations-based methods are not able to provide solutions in a reasonable time,
since the former do not provide easily interpretable results and are less robust to noise.
It is important to note that efficiency of optimization-based methods is influenced not only by the scale of the problem (number of traces under analysis) but also by the shape of RPCs execution time distributions.
This aspect is suggested by the fact that execution times in both optimization-based methods are significantly higher in noised experiments. 
The reason of this behavior can be explained by the fact that different shapes of RPCs execution time distribution
trigger different performance behaviors in optimization-based methods.
For example, branch and bound can less frequently prune branches since bound conditions are not verified,
or distributions with multiple modes can trigger a time-comsuming precomputation in our approach.

Obviously, if the analysis of requests allows to efficiently compute clusters with optimization-based methods,
our proposed approach should be preferred since it outperforms the state-of-art baseline
both in terms of effectiveness and efficiency
(+179\% faster in normal experiments and +175\% faster in noised experiments).
Therefore, we positively answer also to RQ4.

For sake of completeness, all results of both normal and noised experiments are publicly available in \cite{cortellessa_2019}.

%% file: table.tex
\begin{table*}[]
\small
\begin{tabular}{c|c|c|c|c|c|c|c|c|c|c|c|c|c|c|c|}
\cline{2-16}
                         & \multicolumn{3}{c|}{Genetic} & \multicolumn{3}{c|}{Branch and bound} & \multicolumn{3}{c|}{Kmeans}  & \multicolumn{3}{c|}{Hierarchical} & \multicolumn{3}{c|}{Mean shift} \\ \hline
\multicolumn{1}{|c|}{ID} & F-score & Prec & Recall & F-score    & Prec    & Recall    & F-score & Prec & Recall & F-score   & Prec   & Recall  & F-score  & Prec  & Recall  \\ \hline
\multicolumn{1}{|c|}{0}  & 0.844   & 1.0       & 0.73   & 0.743      & 0.741        & 0.745     & 0.678   & 1.0       & 0.513  & 0.678     & 1.0         & 0.513   & 0.349    & 1.0        & 0.211   \\ \hline
\multicolumn{1}{|c|}{1}  & 0.987   & 1.0       & 0.974  & 0.979      & 1.0          & 0.959     & 0.716   & 1.0       & 0.557  & 0.716     & 1.0         & 0.557   & 0.352    & 1.0        & 0.214   \\ \hline
\multicolumn{1}{|c|}{2}  & 0.855   & 0.986     & 0.754  & 0.714      & 1.0          & 0.555     & 0.716   & 1.0       & 0.558  & 0.716     & 1.0         & 0.558   & 0.341    & 1.0        & 0.205   \\ \hline
\multicolumn{1}{|c|}{3}  & 0.855   & 1.0       & 0.747  & 0.862      & 1.0          & 0.758     & 0.961   & 0.925     & 1.0    & 0.956     & 0.92        & 0.995   & 0.272    & 1.0        & 0.157   \\ \hline
\multicolumn{1}{|c|}{4}  & 0.992   & 1.0       & 0.984  & 0.769      & 0.765        & 0.773     & 0.722   & 1.0       & 0.565  & 0.722     & 1.0         & 0.565   & 0.272    & 1.0        & 0.158   \\ \hline
\multicolumn{1}{|c|}{5}  & 0.951   & 1.0       & 0.907  & 0.879      & 1.0          & 0.784     & 0.708   & 0.964     & 0.56   & 0.71      & 0.956       & 0.565   & 0.364    & 1.0        & 0.223   \\ \hline
\multicolumn{1}{|c|}{6}  & 0.992   & 1.0       & 0.984  & 0.888      & 0.799        & 1.0       & 0.681   & 1.0       & 0.516  & 0.681     & 1.0         & 0.516   & 0.43     & 1.0        & 0.274   \\ \hline
\multicolumn{1}{|c|}{7}  & 0.87    & 1.0       & 0.77   & 0.849      & 1.0          & 0.738     & 0.68    & 0.98      & 0.521  & 0.683     & 0.99        & 0.521   & 0.273    & 1.0        & 0.158   \\ \hline
\multicolumn{1}{|c|}{8}  & 0.997   & 1.0       & 0.995  & 0.847      & 1.0          & 0.734     & 0.696   & 1.0       & 0.534  & 0.692     & 1.0         & 0.529   & 0.317    & 1.0        & 0.188   \\ \hline
\multicolumn{1}{|c|}{9}  & 0.904   & 0.976     & 0.843  & 0.873      & 0.974        & 0.791     & 0.733   & 1.0       & 0.579  & 0.731     & 0.991       & 0.579   & 0.468    & 1.0        & 0.305   \\ \hline
\multicolumn{1}{|c|}{10} & 0.837   & 0.993     & 0.724  & 0.871      & 1.0          & 0.771     & 0.696   & 1.0       & 0.534  & 0.696     & 1.0         & 0.534   & 0.353    & 1.0        & 0.215   \\ \hline
\multicolumn{1}{|c|}{11} & 0.992   & 1.0       & 0.984  & 0.867      & 1.0          & 0.766     & 0.692   & 0.98      & 0.535  & 0.692     & 0.98        & 0.535   & 0.26     & 1.0        & 0.15    \\ \hline
\multicolumn{1}{|c|}{12} & 0.984   & 1.0       & 0.969  & 1.0        & 1.0          & 1.0       & 0.982   & 0.965     & 1.0    & 0.982     & 0.965       & 1.0     & 0.374    & 1.0        & 0.23    \\ \hline
\multicolumn{1}{|c|}{13} & 0.985   & 0.975     & 0.995  & 0.987      & 1.0          & 0.974     & 0.696   & 1.0       & 0.534  & 0.696     & 1.0         & 0.534   & 0.415    & 1.0        & 0.262   \\ \hline
\multicolumn{1}{|c|}{14} & 0.815   & 0.951     & 0.714  & 0.619      & 0.978        & 0.453     & 0.706   & 0.939     & 0.565  & 0.706     & 0.939       & 0.565   & 0.395    & 0.923      & 0.251   \\ \hline
\multicolumn{1}{|c|}{15} & 0.992   & 1.0       & 0.984  & 1.0        & 1.0          & 1.0       & 0.733   & 1.0       & 0.579  & 0.733     & 1.0         & 0.579   & 0.43     & 1.0        & 0.274   \\ \hline
\multicolumn{1}{|c|}{16} & 0.926   & 1.0       & 0.862  & 0.676      & 1.0          & 0.51      & 0.7     & 1.0       & 0.538  & 0.696     & 1.0         & 0.533   & 0.259    & 1.0        & 0.149   \\ \hline
\multicolumn{1}{|c|}{17} & 0.834   & 1.0       & 0.716  & 0.838      & 1.0          & 0.721     & 0.68    & 1.0       & 0.515  & 0.68      & 1.0         & 0.515   & 0.438    & 1.0        & 0.281   \\ \hline
\multicolumn{1}{|c|}{18} & 0.828   & 0.953     & 0.732  & 0.663      & 0.695        & 0.634     & 0.7     & 0.964     & 0.549  & 0.7       & 0.964       & 0.549   & 0.429    & 0.981      & 0.275   \\ \hline
\multicolumn{1}{|c|}{19} & 0.734   & 1.0       & 0.58   & 0.844      & 1.0          & 0.731     & 0.669   & 0.99      & 0.505  & 0.671     & 1.0         & 0.505   & 0.38     & 1.0        & 0.234   \\ \hline
\multicolumn{1}{|c|}{20} & 0.997   & 1.0       & 0.995  & 1.0        & 1.0          & 1.0       & 0.683   & 1.0       & 0.518  & 0.683     & 1.0         & 0.518   & 0.398    & 1.0        & 0.249   \\ \hline
\multicolumn{1}{|c|}{21} & 0.845   & 1.0       & 0.732  & 0.849      & 1.0          & 0.737     & 0.669   & 1.0       & 0.503  & 0.669     & 1.0         & 0.503   & 0.292    & 1.0        & 0.171   \\ \hline
\multicolumn{1}{|c|}{22} & 0.878   & 1.0       & 0.782  & 0.875      & 1.0          & 0.777     & 0.689   & 1.0       & 0.526  & 0.689     & 1.0         & 0.526   & 0.407    & 1.0        & 0.255   \\ \hline
\multicolumn{1}{|c|}{23} & 0.886   & 1.0       & 0.795  & 0.832      & 1.0          & 0.713     & 0.7     & 0.972     & 0.546  & 0.707     & 1.0         & 0.546   & 0.403    & 1.0        & 0.253   \\ \hline
\multicolumn{1}{|c|}{24} & 0.853   & 0.993     & 0.747  & 0.855      & 1.0          & 0.747     & 0.691   & 0.99      & 0.531  & 0.691     & 0.99        & 0.531   & 0.318    & 1.0        & 0.189   \\ \hline
\multicolumn{1}{|c|}{25} & 1.0     & 1.0       & 1.0    & 0.896      & 0.811        & 1.0       & 0.703   & 1.0       & 0.543  & 0.703     & 1.0         & 0.543   & 0.291    & 1.0        & 0.17    \\ \hline
\multicolumn{1}{|c|}{26} & 0.852   & 1.0       & 0.742  & 0.852      & 1.0          & 0.742     & 0.688   & 1.0       & 0.524  & 0.688     & 1.0         & 0.524   & 0.418    & 1.0        & 0.265   \\ \hline
\multicolumn{1}{|c|}{27} & 0.995   & 1.0       & 0.989  & 0.887      & 0.987        & 0.805     & 0.674   & 1.0       & 0.508  & 0.674     & 1.0         & 0.508   & 0.371    & 1.0        & 0.228   \\ \hline
\multicolumn{1}{|c|}{28} & 0.955   & 0.923     & 0.99   & 0.73       & 0.829        & 0.653     & 0.714   & 1.0       & 0.555  & 0.709     & 1.0         & 0.55    & 0.585    & 1.0        & 0.414   \\ \hline
\multicolumn{1}{|c|}{29} & 0.686   & 1.0       & 0.522  & 0.686      & 1.0          & 0.522     & 0.72    & 1.0       & 0.562  & 0.724     & 1.0         & 0.568   & 0.263    & 1.0        & 0.151   \\ \hline
\end{tabular}
\caption{Detailed results of noised experiments}
\label{table:results}
\end{table*}

%% file: onestepahead.tex
\section{One step ahead}\label{sec:step}

Often, other information than RPC execution time
are related to performance degradation (e,g. http header, RPC execution node, resource consumption, request size, etc.). 
This information may be often critical to debug performance issues.
In this section, we describe how our problem can be easily extended to trace attributes other than RPC execution time.
Typically, a trace attribute can either be \emph{categorical} or \emph{continuous}.
A \emph{categorical attribute} can take one of a limited and usually fixed number of possible values (e.g. request country, request language, RPC execution node, etc.),
while a \emph{continuous attribute} can take any value in a particular interval (e.g. request size, RPC response time, etc.).

The problem formalized in Section \ref{sec:formaldef} can be generalized as follows:

\begin{definition}
A trace $r$ is an ordered sequence of attributes $r=(a_0, a_1, ..., a_m, L)$,
where $a_j$ denote the observed value of a traced attribute $j$,
and $L$ is the observed latency of the request.
\end{definition}

\begin{definition}
A trace attribute $j$ can be either \emph{categorical} or \emph{continuous}.
A categorical attribute can take a value $a_j$ in a finite set $A_j$, $a_j\in A_j$.
A continuous attribute can take value in a continuous interval $a_j\in [a_j^{min}, a_j^{max})$.
\end{definition}

According to the attribute type, a condition can be either categorical or continuous.

\begin{definition}
A categorical condition is a pair $c=\langle j, v \rangle$,
where $j$ is a categorical attribute and $v\in A_j$.
A request $r=(..., a_j, ...)$ satisfies $c$, denoted as $r\vartriangleleft  c$ , if and only if $a_j=v$.
\end{definition}

\begin{definition}
A continuous condition $c$ is a triple $c=\langle j, v_{min}, v_{max} \rangle$,
where $j$ is a continuous attribute, and $[v_{min}, v_{max})$ represent a sub-interval of the interval $[a_j^{min}, a_j^{max})$.
A request $r=(..., a_j, ...)$ satisfies $c$, denoted as $r\vartriangleleft  c$ , if:
\begin{equation*}
 v_{min}\leq a_j < v_{max}
 \end{equation*}

\end{definition}

Definition \ref{def:pattern} of pattern still holds.

Obviously, this generalization will impact the genetic representation of a solution,
as well genetic variation operators and fitness optimization.

%% file: relatedwork.tex
\section{Related work}\label{sec:relatedwork}

Workflow-centric tracing research area \cite{Sambasivan2016} is certainly related to our work.
Dapper \cite{Sigelman2010} was the pioneering work in this space,
Canopy \cite{Kaldor2017} processes traces in real-time, derives user-specified features, and outputs performance datasets that aggregate across billions of requests. 
Pivot Tracing \cite{Mace2016} gives users the ability to define traced metrics at runtime,
even when crossing component or machine boundaries.

Interesting work in the area of supporting performance diagnosis has appeared in the last few years.
The study of Sambavisan et al. \cite{Sambasivan2013} compares three well-known visualization approaches in the context of results presentation of an automated performance root cause analysis approach \cite{Sambasivan2011}.
Malik et al. \cite{Malik2010} use performance counter data of a load test to craft performance signatures and use them to pinpoint the subsystems responsible for the performance violations.
In \cite{Malik2013} four machine learning approaches are presented and evaluated to help performance analysts to more effectively compare load tests in order to detect performance deviations which may lead to SLA violations, and to provide them with a smaller and manageable set of important performance counters.
StackMine \cite{Han2012} also aims to improve performance debugging by reducing the scope of the callstack traces analysis.

The closest work to ours is the approach by Krushevskaja and Sandler \cite{Krushevskaja2013}, which we have in fact used in our experimentation for sake of comparison.
They introduced multi-dimensional f-measure, that is an information retrieval metric helping to identify the best values of monitored attributes for a given latency interval.
They also proposes algorithms that use this metric not only for a fixed latency interval (branch and bound and forward feature selection), but also to explain the entire range of service latency by segmenting it into smaller intervals.